# A toolkit of dilemmas: Beyond debiasing and fairness formulas for responsible AI/ML

Andrés Domínguez Hernández
*Computer Science*
*University of Bristol*
Bristol, UK
andres.dominguez@bristol.ac.uk

Vassilis Galanos
*Science, Technology and Innovation Studies*
*University of Edinburgh*
Edinburgh, UK
vassilis.galanos@ed.ac.uk

*Abstract*—Approaches to fair and ethical AI have recently fell under the scrutiny of the emerging, chiefly qualitative, field of critical data studies, placing emphasis on the lack of sensitivity to context and complex social phenomena of such interventions. We employ some of these lessons to introduce a tripartite decision-making toolkit, informed by dilemmas encountered in the pursuit of responsible AI/ML. These are: (a) the opportunity dilemma between the availability of data shaping problem statements versus problem statements shaping data collection and processing; (b) the scale dilemma between scalability and contextualizability; and (c) the epistemic dilemma between the pragmatic technical objectivism and the reflexive relativism in acknowledging the social. This paper advocates for a situated reasoning and creative engagement with the dilemmas surrounding responsible algorithmic/data-driven systems, and going beyond the formulaic bias elimination and ethics operationalization narratives found in the fair-AI literature.

## I. Introduction

### A. Rationale, Background, and Outline

A wealth of recent critical studies has evidenced how machine learning (ML) algorithmic systems trained on biased datasets, questionable scientific grounds and without attention to context can have uneven impacts on different demographics and disproportionally harm marginalized communities [1]–[3]. Researchers within computer science and data science have advanced frameworks of fairness, risk scoring and explainability in response to the ethical concerns and critiques from social science [4]–[7]. The focus of this work has been largely on developing fair or unbiased models, operationalizing ethical frameworks and improving documentation and intelligibility. But such a calculus-oriented approach might find itself 'trapped' with conflicting definitions of fairness and disagreements about what the morally right outcomes are or should be, due to poor assessments of real-world conditions or developers' inexperience with the contexts where algorithms have an impact [8], [9]. AI and ML development has remained removed from reflections on how the epistemologies, motivations and politics underpinning the development of algorithmic systems influence their social impacts, in particular when it comes to priority-setting in AI development [10]. We contend that existing framings of responsible AI/ML still need to address a series of issues which previous theories and studies of knowledge production have explored, especially in the field of Science and Technology Studies (STS).

In this paper we engage closely with the ambivalence, ambiguity and indeterminacy that are inherent to many areas of application of algorithmic systems [9], and wish to offer a toolkit for informing responsible decision making. We propose looking at wicked challenges as opportunities to negotiate ethical outcomes through a situated and creative pondering of dilemmas. We put forward three quandaries arising in the pursuit of responsible development of algorithmic systems, further analyzed and problematized in this paper's second section: (a) *The opportunity dilemma*, or what comes first, datasets or problem statement? (b) *The scale dilemma*, or the tension between cost-efficiency pressures to scale up versus the demand to situate and contextualize based on social specificities; and (c) *The epistemic dilemma* of locating responsible interventions between naïve-objectivist measurements or impractical relativist assessments. Our proposal stems from our long-term engagement with AI/ML practitioners as members of collaborative research projects and draws from the body of critical scholarship in STS and allied fields. We offer a theoretical framework based on overlapping themes within different practical contexts in AI/ML design. We suggest that this framework's dilemmas should be components of discussions and reports addressing AI/ML applications' design limitations towards efforts in making the process more transparent and responsible. Our provocation wishes to invite critical intuition and interdisciplinary dialogue in relation to contemporary crises of legitimacy, truth and trust with data-driven technology and, more broadly, the epistemological footings and hyped claims of algorithmic systems.

## II. Theoretical contribution: The Three Dilemmas for AI Researchers

In this section we outline the three dilemmas arising in contemporary discussions about responsible innovation within computational and social sciences. We suggest that although they appear as problematic and unsolvable, they should not be considered as obstacles, but as parameters of sensitization for AI/ML practitioners. Similarly, we note that these dilemmas are not exhaustive and will not present themselves equally in all scenarios: resolutions in certain areas might be more straightforward than in others. One possible application in the context of responsible innovation practices [11]–[13] would be to audit the practitioners' or designers' responses and rationalization for selecting one option in favor of another, or where that may be the case, introducing innovative (including non-technical) methods to resolve the dilemma. Table 1 summarizes the three proposed dilemmas and their options, briefly outlining their advantages and disadvantages. Such dilemmas should not be perceived as clear-cut either-or choices, but as part of paired of antinomies, leading into novel knowledge representations, as recently argued in the context of bridging knowledge management with data labelling practices in a process of dialectical synthesis, where two apparent contradictions act as opportunities for social change [11].

*A. The Opportunity Dilemma: What goes first? Data or Problem Statements?*

The first dilemma we identify has to do with what drives innovation and priorities in AI development. Do problem statements shape the definition and preparation of the datasets needed? Or the opportunistic availability of data, and inspecting the available affordances helps to shape the problem to be tackled by AI? A great deal of research in AI is driven by technological solutionism based on what is technically, economically and even politically feasible. The availability of data is a key criterion for the development of applications. If data availability precedes the formulation of problems to be tackled by AI, this might be justified mainly by opportunity whereby the problematization of constraints and path dependencies engendered by existing data become second order concerns. Consider for example the implications of using legacy databases based on binary sex and gender classifications for automating decision making concerning the rights of trans gender and non-binary individuals, both in terms of sexual/gender recognition claimed based on facial recognition datasets [12], or body scanning x-ray datasets [13]. Such *data first* approaches could lead to harmful mismatches of body characteristics conventionally associated with particular gender identities (here, consider the implications for breast cancer ML screening [14]). The alternative would be to start from an understanding of the needs and vulnerabilities of social groups, in this case the LGBTQI+ community. This might call for the construction of new datasets that are reflective of such diversities –an approach which requires more resources and political persuasion to secure those resources. In grappling with this dilemma, we should ask whether the applications being developed respond to solely a logic of opportunity and technology push; to address pressing social needs; or both. Moreover, this dilemma requires us to grapple with questions around routine and serendipity in the process of dataset labelling and clustering. As qualitative studies have shown, everyday labelling work is seldom a well-informed practice performed by technical experts with prior knowledge about the social significance of the data aggregates being clustered and labelled [15], [16]. Instead, labelling is often a precarious and mundane process which involves, for example, workers investing more time in creating ideal types of clustered data and less in reflecting on the representational truthfulness of a given label. Data work is likely to happen under the pressure of meeting deadlines and is susceptible to the interpretative flexibility of numerical outputs based on the data scientists' understanding of a given cluster's symbolic value. In other words, not every data scientist can know about the particularities of a given cluster or dataset's social context, especially when the numerical representation of a cluster is alienating the labeler from the situation in which data have been collected.

*B. The Scale Dilemma: Generalize or Contextualize?*

Two competing views on data exist in discussions about responsible AI/ML: more is more and less is more. Some definitions of fairness suggest more, and better-quality data will mitigate algorithmic bias and injustice. Such is the preferred approach of conventional data science, which aims to satisfy the cost and moral exigencies of reducing harm while increasing financial return of more accurate findings arising from even greater volumes of data. However, the *more-is-more* argument is less concerned with issues where "too much data" might have harmful effects to those involved in annotation work as shown in several critical case studies [1], [2]. Too much data could also be technically detrimental. For example, in medical imaging applications of ML, a recent study suggested that "more samples may yield poorer performance" as they obscure small, yet important, structural changes, instabilities, and nearly undetectable perturbations based on a variety of social factors, becoming easily dismissed in large-scale pattern analysis, thus the authors' calling for more context-based testing of such instability phenomena [17]. The other approach is influenced by the social exigencies to situate datasets within their given contexts in order to serve the needs and requirements of specific and underrepresented groups impacted by algorithmic systems [18]. The latter *less-is-more* approach suggests that situated, small scale, qualitative and participatory approaches are necessary and that non-technical solutions should be part of the repertoire of responsible innovation. However, scaling down and prioritizing contextual accuracy through more situated approaches might face funding barriers, risks of participation washing or causing epistemic burden to marginalized groups and might be difficult to scale up [19], [20]. Moreover, an ultra-specific dataset defeats the very purpose of modelling, harassing the map-territory relationship, which is why context-based datasets and models are expected to have at least some mild degree of generalizability. In the field of ML, this is known at least since 1980, as the need for acknowledged and combined use of "prior knowledge, biases, and observation in guiding the learning process" [21], which feeds back into the process of the first dilemma. In pondering this dilemma researchers need to revisit the pervasive and taken-for-granted dictum of scaling up, and consider alternatives for broader social alignment for specific purposes [22] between data practitioners, data generators, and data owners.

*C. The Epistemic Dilemma: Objectivism or Relativism?*

The third dilemma we have identified lies on the deeper epistemological standpoint preferred by the designer or data scientist who subscribes to a responsible intervention approach to algorithmic design rooted in positivist, techno-centric values of cost-efficiency, performance and objectivity [23]. The implicit pragmatism of data science suggests that the method is appropriate and to a large extent objective, at least as long as there are defensible ways to justifying the truthfulness of a dataset. The social scientific critique, on the other hand, would problematize objectivity as a naïve ideal by highlighting the contingent character of any scientific endeavor to social, cultural, political, and material factors. However, this critique is easy to fall into the traps of endless debate and "mind games," which is often perceived as halting the pragmatic need for socially positive outcomes or product delivery [24]–[26]. As much as objectivist science is proven to be not an authority independent of social context, relativist science risks reducing practical calls to action and being weaponized by anti-science actors, as per previous lessons from the "science wars" debate [27]. Exits to the tension between pragmatic-yet-naïve objectivism and reflexive-yet-futile relativism are found in critical disciplinary reformulations, open to techno-social change, such as Birhane's proposition to shift from rational formalisms within computational sciences to a relational ethics approach [28], or the growing calls to import reflexivity and qualitative methods into the design of algorithmic systems [29]–[31].

TABLE I: A Toolkit Of Dilemmas

| A Toolkit of Dilemmas | | |
|---|---|---|
| *Opportunity Dilemma* | *Scale Dilemma* | *Epistemic Dilemma* |
| *Datasets followed by problem statement* | *Generalizable Datasets* | *Objectivist Measurements* |
| Advantage: stable decision making practicing<br><br>Disadvantage: constantly updated data and broader contexts/semantics, path dependencies of static legacy datasets influencing problematization and perpetuating injustices | Advantage: Cost-efficiency. Improved recommendations based on variety<br><br>Disadvantage: disproportionate representation of (and harm to) marginalized groups and context-specific needs; explainability concerns, ease to employ datasets towards desired outcomes. Not everything should be scaled | Advantage: highly utilitarian precision metrics<br><br>Disadvantage: "computer said so" effects; performativity of misleading metrics of performance; Harm to minoritized, unrepresented communities |
| *Problem statements followed by datasets* | *Context-specific Datasets* | *Relativist critique* |
| Advantage: well-informed responsible approach, applications aligned with legitimate social needs, technological approach is not a pre-requisite<br><br>Disadvantage: potential lack of appropriate datasets, economic unfeasibility | Advantage: Accountability for specificity of needs and particularities of social semantics<br><br>Disadvantage: Funding barriers, epistemic burden, lack of generalizable utility | Advantage: Acknowledgement of social richness, power and awareness of multiple challenges. Flagging of harmful generalizations<br><br>Disadvantage: High-level, loose and vague recommendations. Unclear applicability. Pragmatic futility in high-speed technical deployment |

### III. CONCLUSIONS AND FURTHER WORK

The last decade has seen a burgeoning of engaged and detailed work from various fronts within the social sciences into the contestability of sources of "truth" employed for ground-truth datasets, the politics of algorithms, the invisible material and human cost of curation processes, the performativity of algorithmic metrics and AI future imaginaries, and an assessment of individual/institutional motives as relating to the social and/or user-specific benefit [9], [30], [32]. Yet, there remains a huge gap to integrate this wealth of social scientific inquiry into AI/ML development, regulation and education. To contribute to tackling this challenge, we have hereby presented an array of dilemmas arising in the pursuit of responsible, fair and just development of algorithmic systems. While the framing of dilemmas usually evokes a sense of analytic paralysis or are resolved quickly through loosely justified formulas of acceptable balance and trade-offs, we recommend deep and continual interrogation of them during discussions about responsible AI/ML. The Zen Buddhist logic of introducing unsolvable paradoxes as riddles (koans) to meditating students acts as an opportunity for the latter to overcome the paradox, thus reaching enlightenment. Such approaches of treating everyday dilemmas as meditation paradoxes have found fertile operation in clinical practice [33] – why not in ML practice too? Does the designer wish to define the problem based on existing datasets or find datasets to match existing problem definitions? Do we want generalizable, cost-effective data, or specialized, context-specific ones? Do we prefer the naiveté of objectivism or the futility of relativism? The best answers to these questions might lie beyond simple binaries, if we approach them as commensurable opportunities for nuanced new options, akin to dialectical synthesis [11]. Such interrogations, we hope, invite creative and meaningful debate, and seek to avoid forgone conclusions and blind acceptance of dominant discourses of inevitability and solutionism in AI/ML.

ACKNOWLEDGMENT

We declare that no funding has been received for authoring this paper and no conflicts of interest have been identified during the process. ADH wishes to thank REPHRAIN under UKRI grant: EP/V011189/1 for support during the writing of this paper.